\documentclass[%
 reprint,amsmath,amssymb,aps]{revtex4-1}
\usepackage{graphicx}
\usepackage{dcolumn}
\usepackage{bm}
\usepackage{braket}
\usepackage{epstopdf}
\usepackage[sort&compress]{natbib}
\begin{document}
\title{Atom-assisted quadrature squeezing of a mechanical oscillator inside a dispersive cavity}
\author{Anil Kumar Chauhan}
 \email{anilkm@iitrpr.ac.in}
\author{Asoka Biswas}%
\affiliation{%
Department of Physics, Indian Institute of Technology Ropar, Rupnagar, Punjab - 140001, India
}%
\date{\today}%

\begin{abstract}
We present a hybrid optomechanical scheme to achieve dynamical squeezing of position quadrature of a mesoscopic mechanical oscillator, that can be externally controlled by classical fields. A membrane-in-the-middle set up is employed, in which an atom in $\Lambda$ configuration is considered to be trapped on either side of the membrane inside the cavity. We show that a considerable amount of squeezing (beyond the 3 dB limit) can be achieved that is not affected by the decay of the cavity and the spontaneous emission of the atom. Squeezing depends upon the initial preparation of atomic states. Further, a strong effective coupling between the atom and the oscillator can be attained by using large control fields that pump the atom and the cavity. Effect of thermal phononic bath on squeezing is studied in terms of the squeezing spectrum. The results are supported by the detailed analytical calculations.
\end{abstract}

\maketitle

\section{\label{sec:i}Introduction\protect\\   }
Detection of quantum effects in mesoscopic harmonic oscillator (MHO) has been into the focus of study for quite a long time. Such an oscillator is composed of a few billion atoms and therefore can be considered as a system of classical nature. Interestingly, at low temperature, it can be driven into a quantum state, e.g., a superposition of separated position eigenfunctions.
Enormous attempts have been taken to reach the quantum regime in the MHO. A central thrust of this effort has been the development of ultra-sensitive displacement measurement techniques. The measurement of position of an oscillator of mass $m$ in quantum regime is however limited by the standard quantum limit (SQL; $(\Delta x)_{\rm SQL}=\sqrt{\hbar/2m\omega_m}$), arising due to the intrinsic zero-point fluctuation, where $\omega_m$ is the natural frequency of the oscillator. In addition, the oscillator also gets perturbed by the measurement device itself in quantum regime, leading to the back-action of the oscillator onto the measurement device. This increases the minimum limit of achievable uncertainty of the position to $\sqrt{2}(\Delta x)_{\rm SQL}$. Till date, the best possible uncertainty that could have been achieved is $\sim 4(\Delta x)_{\rm SQL}$ \cite{lahaye} in a nanomechanical oscillator ($\omega_m=1.35$ GHz), coupled to a single-electron transistor, while uncertainties $\sim 100 (\Delta x)_{\rm SQL}$ \cite{knobel} and $\sim 30(\Delta x)_{\rm SQL}$ \cite{regal} are also reported in lower-frequency oscillators.
Back-action evading techniques with ideally infinite measurement precision have been proposed \cite{clerk-njp} and demonstrated \cite{hertzberg-clerk} to achieve up to $\sim 1.3(\Delta x)_{\rm SQL}$; however it yields less information and requires the measurement to be much faster than the relaxation time of the oscillator.

Measurement of position below SQL has seen a growing interest in recent times in the context of cavity optomechanical systems \cite{rmp}. Generating non-classical states like position-squeezed states in this system can lead to $(\Delta x)< (\Delta x)_{\rm SQL}$. Such a system consists of a single mode Febry-Perot cavity with one movable end mirror, in which the coupling between the cavity mode and the mechanical mode of the mirror is created due to the radiation pressure force.  It has been considered as a test platform to explore possibilities of squeezing in mesoscopic oscillators.
The radiation pressure force makes the coupling between the two modes, linear in $x$, the displacement of the mirror from its equilibrium position. Several proposals have been made to achieve position squeezing in such systems. For example, the position squeezing can be obtained by pumping the cavity by a squeezed light source and thereafter transferring this squeezing to the oscillator mode through a state transfer protocol \cite{jahne2009}. A two-mode cavity can be made equivalent to an engineered reservoir
that can lead to squeezing of the oscillator via feedback \cite{feedback}. It is also shown that short pulses can be
used to obtain mechanical squeezing in optical microcavity \cite{vanner-pnas}. Such methods however require either a continuous source of squeezed light and high transfer efficiency at quantum level or short pulses and thereby are not the most sought-after methods for squeezing.

A natural way of obtaining quadrature squeezing dynamically is to use a Hamiltonian that is quadratic in position quadrature $X=(b+b^\dag)/2$  or momentum quadrature $P=(b-b^\dag)/2i$, where $b$ is the annihilation operator of the quantized oscillator. e.g.,  the Hamiltonian $H=\chi(b^2+b^{\dag 2})$, $\chi$ being equivalent to the squeezing parameter, that is similar to that of a degenerate parametric amplifier \cite{scully-book}. Position squeezing in ground state of the oscillator in presence of back-action would refer to $(\Delta X)^2<(\Delta X)^2_{\rm SQL}=1/2$. It was shown in \cite{Thompson2008} that if a mechanical oscillator is suspended inside a cavity (with both the mirrors fixed) at a position where frequency $\omega_c$ of the cavity sees a node or anti-node (i.e., $\partial \omega_c/\partial x=0$, a ``membrane-in-the-middle" set up), the coupling becomes quadratic in the displacement of the oscillator. In such a system, squeezing can be obtained through a unitary evolution. Driving the cavity with two laser beams, whose frequencies are detuned to either side the
cavity resonance by an amount equal to the mechanical frequency in such a
set-up, one can also obtain \cite{nunnenkamp} a quadratic Hamiltonian. The squeezing property of a quadratic Hamiltonian is discussed in details in \cite{shi-bhatta} in the context of cavity optomechanical systems. It is shown that to obtain a
large squeezing, one requires a large number of average thermal photons and a proper
conditional measurement of photon numbers in the cavity. This is however limited by the cavity decay, as large number of photons are prone to faster decay out of the cavity and it can lead to degradation of squeezing. To combat this dissipation, alternative methods have also been proposed, that require applying three coherent fields \cite{roque} or periodic intense pulses \cite{asjad}, as commonly used in dynamical decoupling techniques \cite{viola}.

In all the above methods to obtain squeezing, one employs either a passive method such as feedback, or a set of coherent fields or pulses. Further, though one can dynamically achieve squeezing through a quadratic Hamiltonian, the squeezing effect is not pronounced, because, the optical cavity decays much faster than the oscillator.
In this paper, we propose a scheme to obtain dynamical squeezing, in which the effect of cavity decay is eliminated. This would be possible, if the cavity mode interacts dispersively with the system, while squeezing in the oscillator is governed by another auxiliary system, say, an atom. Specifically, we consider an atom-cavity-oscillator hybrid system, in which a coherently driven atom is coupled to the mechanical oscillator via their common coupling to the cavity mode in the membrane-in-the-middle set up. The cavity mode is adiabatically eliminated from the interaction and thus the cavity decay can be avoided. Further, we choose the low-lying energy levels of the atom, which are immune to spontaneous emission.  In this way, the steady state squeezing becomes independent of any mode of decay in the system. The interesting features of this model are: (a) The control fields, that drive the atomic transition and the cavity mode, control the degree of squeezing in the position quadrature of the oscillator. (b) The squeezing parameter depends upon the initial state of the atom. Note that cavity optomechanics mediated by a two-level system has also been proposed in \cite{nature2015}, in which a Josephson-junction qubit strongly interacts directly with both a microwave cavity and the micromechanical oscillator. On the contrary, in the present model, the cavity mode, instead of the atom, mediates the coupling between the atom and the oscillator and the intrinsic atom-cavity and oscillator-cavity couplings are weak in nature, that further can be controlled by external pumping fields.

The paper is organized as follows. We describe our hybrid system in Sec.~\ref{s:ii}. We discuss how the effective atom-oscillator Hamiltonian can be obtained via adiabatic elimination of the cavity mode. We also derive the expression of the squeezing in time-domain as well as in frequency domain. We discuss the effect of the thermal bath on squeezing. Results are  discussed in detail in Sec.~\ref{s:iii}, along with comparison with other proposals on squeezing.
We conclude the paper in Sec.~\ref{s:iv}.
-----------------------
\begin{figure}[!h]
$\begin{array}{cc}
    \includegraphics[width=.6\linewidth]{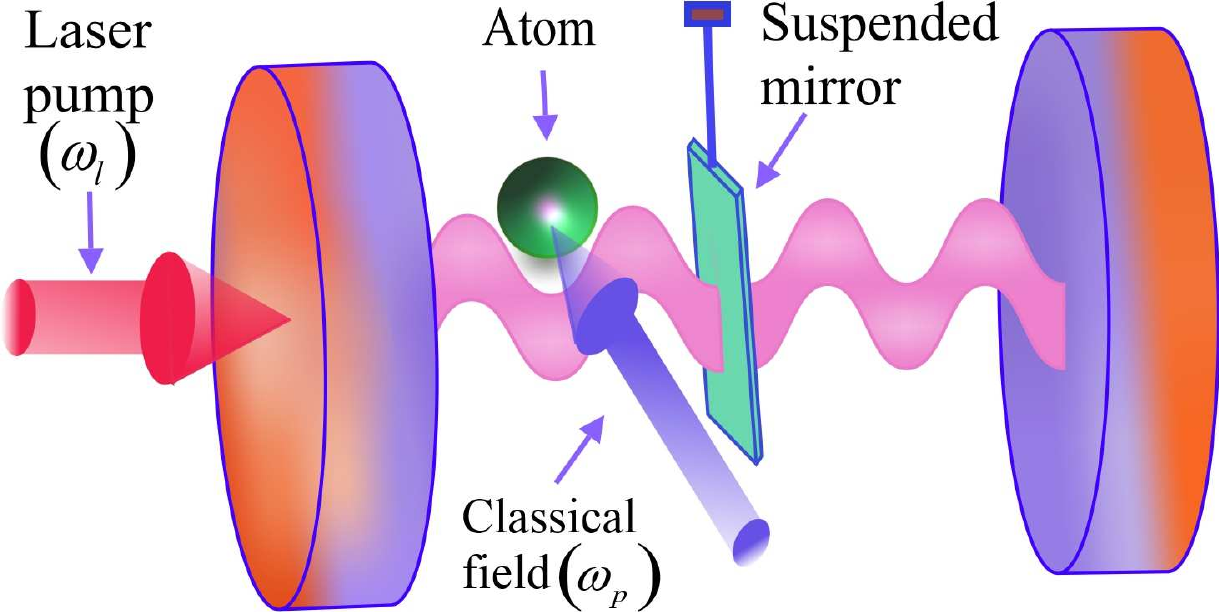}&
    \includegraphics[width=.38\linewidth]{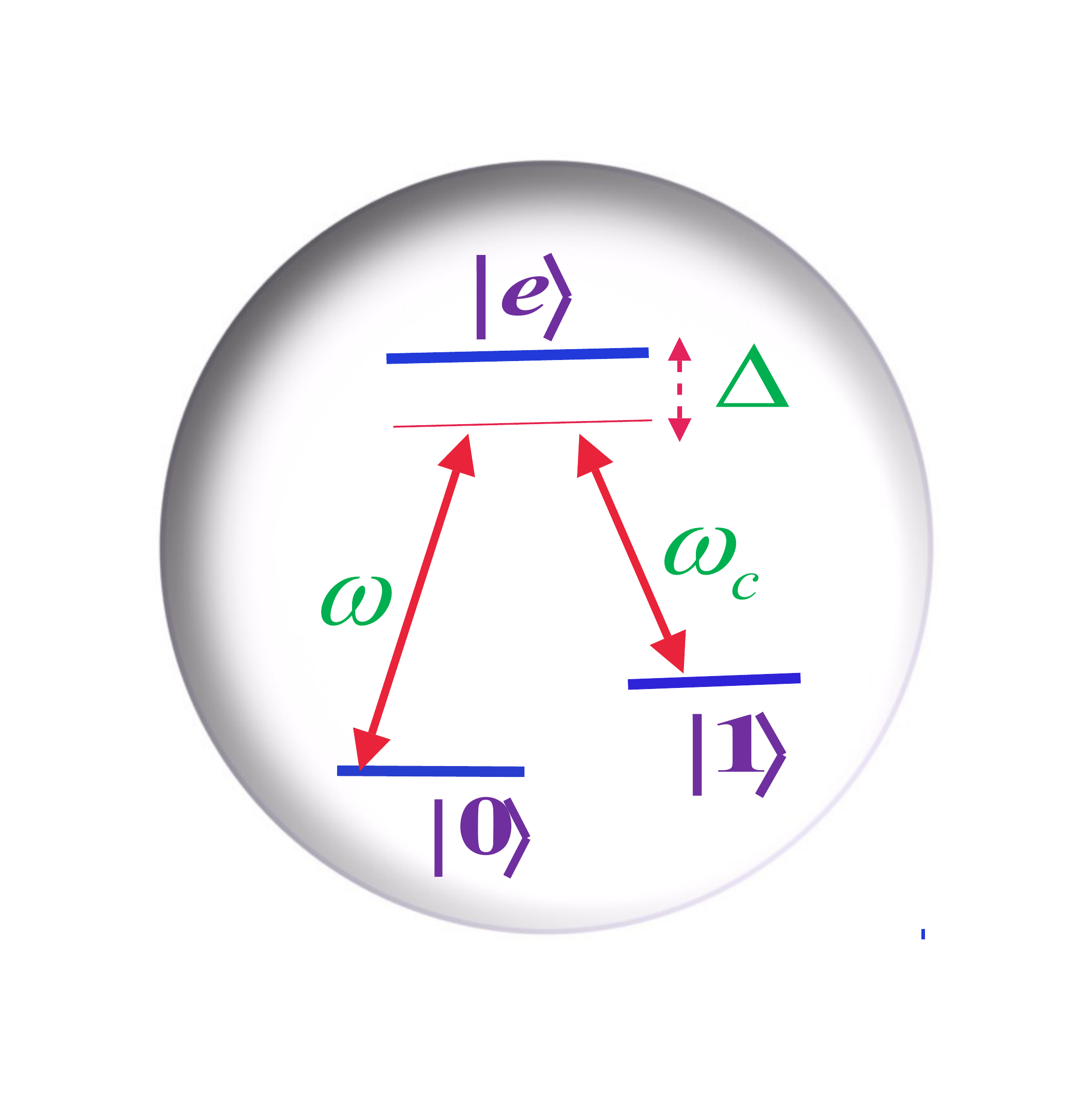}\\(a) & (b)
\end{array}$
 \caption{(a) A membrane-in-the-middle set up, in which an atom is trapped and a mechanical oscillator is suspended inside a driven cavity, (b) Energy level configuration of the atom.}
\end{figure}

\section{A  hybrid Model}\label{s:ii}
We consider a mechanical oscillator suspended inside an optical cavity, that has both the mirrors fixed (the ``membrane-in-the-middle" setup) \cite{Thompson2008}. The dynamics of this system is governed by the following Hamiltonian (in $\hbar=1$ unit):
\begin{equation}
H_1=H_0+H_{\rm cm}+H_{\rm pump}\;,
\end{equation}
where \begin{eqnarray}
H_0&=&\omega_c a^\dag a+\omega_m b^\dag b\;,\nonumber\\
H_{\rm cm}&=&g_2a^\dag a(b+b^\dag)^2\;,\nonumber\\
H_{\rm pump}&=&\epsilon (a^\dag e^{-i\omega_l t}+{\rm h.c.})\;.
\end{eqnarray}
Here $\omega_c$ and $\omega_m$ are the frequencies of the cavity mode $a$ and the mechanical oscillator mode $b$, respectively, $g_2$ defines the coupling between them, and $\epsilon$ is the amplitude of the coherent field of frequency $\omega_l$ that pumps the cavity mode. Note that the $g_2$ is proportional to the second-order derivative of $\omega_c$ with respect to the displacement $x$ of the oscillator from its equilibrium position. This Hamiltonian is quadratic, as $H_{\rm cm}$ is proportional to the second order of the operators $b$ and $b^\dag$ of the oscillator.

In our hybrid model, we consider a single atom with three relevant energy levels $|0\rangle,|1\rangle,|e\rangle$ in $\Lambda$-configuration, that is magneto-optically trapped inside the cavity on either side of the oscillator [see Fig. 1]. The $|0\rangle\leftrightarrow |e\rangle$ transition is driven by a classical control field with frequency $\omega_p$ and the Rabi frequency $\Omega$, while the cavity mode drives the $|1\rangle\leftrightarrow |e\rangle$ transition. The relevant Hamiltonian of the atom-cavity system can be written as
\begin{eqnarray}
H_{ac}&=&\Omega e^{-i\omega_p t}|e\rangle\langle 0|+g_1|1\rangle\langle e|a^\dag +{\rm h.c.}\;,\nonumber\\
H_0^{\rm atom}&=&\omega_{e0}|e\rangle\langle e|+\omega_{10}|1\rangle\langle 1|\;,
\end{eqnarray}
where $g_1$ is the atom-cavity coupling constant and $H_0^{\rm atom}$ is the unperturbed Hamiltonian of the atom.
In the reference frame, rotating with the pumping laser frequency $\omega_l$, the total Hamiltonian $H=H_1+H_{\rm ac}+H_0^{\rm atom}$ of the hybrid system reduces to
\begin{equation}
H^{(1)}=H_0^{(1)}+H_{\rm ac}^{(1)}+H_{\rm cm}+H_{\rm pump}^{(1)}\;,
\end{equation}
where \begin{eqnarray}
H_0^{(1)}&=&\delta a^\dag a+\omega_mb^\dag b+\omega_{e0}|e\rangle\langle e|+\omega_{10}|1\rangle\langle 1|\;,\nonumber\\
H_{\rm ac}^{(1)}&=&(\Omega e^{-i\omega_p t}|e\rangle\langle 0|+g_1|1\rangle \langle e|a^\dag e^{-i\omega_l t}+{\rm h.c.})\;,\nonumber\\
H_{\rm pump}^{(1)}&=&\epsilon (a+a^\dag)\;,
\end{eqnarray}
and $\delta = \omega_c-\omega_l$ is the cavity pump detuning.
Next, in the interaction picture with respect
to the Hamiltonian $H_0^{\rm atom}$, the Hamiltonian $H^{(1)}$ takes the following form:
\begin{equation}
H^{(2)}=H_0^{(2)}+H_{ac}^{(2)}+H_{\rm cm}+H_{\rm pump}^{(1)}\;,
\end{equation}
where \begin{eqnarray}
H_0^{(2)}&=&\delta a^\dag a+\omega_mb^\dag b\;,\nonumber\\
H_{\rm ac}^{(2)}&=&(\Omega e^{i\Delta t}|e\rangle\langle 0|+g_1|1\rangle\langle e|e^{i(\Delta+\delta)t}a^\dag+{\rm h.c.})\;,
\end{eqnarray}
and $\Delta=\omega_{e0}-\omega=\omega_{e1}-\omega_c$ is the common detuning of
the laser field and the cavity mode from the respective one-photon transition.
\subsection{Effective Hamiltonian}
We consider the large detuning limit $\Delta\gg \Omega,g_1$. With this approximation, the
excited state $|e\rangle$ can be eliminated adiabatically \cite{Biswas2004} and the three-level
atom reduces to an effective two-level atom, with the relevant energy-levels
$|0\rangle$ and $|1\rangle$. The Hamiltonian then can be written as
\begin{eqnarray}
H^{(3)}=H_0^{(2)}+H_{ac}^{(3)}+H_{\rm cm}+H_{\rm pump}^{(1)},
\end{eqnarray}
where
\begin{eqnarray}
H_{\rm ac}^{(3)}=-\frac{\Omega g_1}{\Delta}(|0\rangle\langle 0|+|1\rangle\langle 1|a^\dag a)\nonumber\\
-\frac{\Omega g_1}{\Delta}(|0\rangle\langle 1|a+{\rm h.c.})-\delta|1\rangle\langle 1|\;.
\label{hac}
\end{eqnarray}
The first term in (\ref{hac}) above represents the Stark shifts of the ground states of the atom due to its coupling to the control field and the cavity field, while the second term describes the dispersive coupling between the atom and the cavity mode.

The Heisenberg equation of motion of the cavity mode $a$ can be written as
\begin{eqnarray}
\dot{a}=-i[a,H^{(3)}]= -i\left[\delta a+\epsilon -\frac{\Omega g_1}{\Delta}|1\rangle\langle 1|a\right.\nonumber\\
-\left.\frac{\Omega g_1}{\Delta}|1\rangle\langle 0|+g_2(b+b^\dag)^2a\right]\;.
\end{eqnarray}
In the limit, $\delta\gg\frac{\Omega g_1}{\Delta},g_2$, we can adiabatically eliminate the cavity mode $a$ by choosing $\dot{a}\approx 0$. This leads us to the following operator identity: $a\approx \frac{1}{\delta}\left(\frac{\Omega g_1}{\Delta}|1\rangle\langle 0|-\epsilon\right)$. Thus, the final effective Hamiltonian becomes
\begin{equation}
H^{(4)}=\omega_mb^\dag b+\hat{g}(b+b^\dag)^2\;,
\label{finalH}
\end{equation}
where
\begin{equation}
\hat{g}=\frac{g_2}{\delta^2}\left[\left(\frac{\Omega g_1}{\Delta}\right)^2|0\rangle\langle 0|
-\epsilon\frac{\Omega g_1}{\Delta}(|0\rangle\langle 1|+|1\rangle\langle 0|)+\epsilon^2\right]\;.\\
\label{g}
\end{equation}
Clearly, (\ref{finalH}) depends quadratically on the position quadrature (proportional to $b+b^\dag$) of the oscillator. This is the desired form of the Hamiltonian to obtain squeezing through a unitary evolution. This is also equivalent to the Hamiltonian that gives rise to quantum optical spring effect \cite{amit-gsa}, in which decay of the atom or the cavity mode is now effectively eliminated.

Further, $\hat{g}$ defines an atomic operator, indicating that by suitably choosing the initial state of the atom, one can control the squeezing parameter.
This can be further revealed by rewriting $\hat{g}$ in the eigenbasis of its atomic part as
\begin{eqnarray}
\hat{g}=\frac{g_2}{\delta^2}\left[\lambda_1|e_1\rangle\langle e_1|+\lambda_2|e_2\rangle\langle e_2|+\epsilon^2\right]\;,
\end{eqnarray}
where $\lambda_{1,2}=\frac{1}{2}\frac{\Omega g_1}{\Delta}\left[\frac{\Omega g_1}{\Delta}\pm \sqrt{\left(\frac{\Omega g_1}{\Delta}\right)^2+4\epsilon^2}\right]$ and $|e_{1,2}\rangle$ are the corresponding eigenstates. If the atom is prepared in one of these eigenstates $|e_i\rangle$ ($i=1,2$), the effective coupling constant takes the form $g_{\rm eff}=(g_2/\delta^2)[\lambda_i+\epsilon^2]$. Therefore one can obtain desired squeezing by a suitable choice of $\Omega$ and $\epsilon$.
Note that one could also achieve squeezing if the oscillator is parametrically driven so that the coupling constant becomes a sinusoidal function of time. This is usually done in a movable-mirror set-up, in which the frequency of the cavity pump laser is suitably modulated \cite{nunnenkamp,mari}, while in the present hybrid model, one does not require any modulation.

\subsection{Squeezing}
We assume that the oscillator is in thermal equilibrium with the phononic environment at a temperature $T$. The state of the oscillator is described by the density matrix $\rho=\sum_np_n|n\rangle\langle n|$, where $p_n=(1-\exp[-\hbar\omega_m/k_BT])\exp(-n\hbar\omega_m/k_BT)$ is the probability that the oscillator is in the phonon number state $|n\rangle$ and $k_B$ is the Boltzmann constant. To identify squeezing in position, we calculate the time-dependent uncertainty of the relevant quadrature $X=(b+b^\dag)/2$. In Heisenberg picture, the operator $b$ evolves as
\begin{equation}
b(t)= \exp[iH^{(4)}t]b\exp[-iH^{(4)}t]=rb(0)+sb^\dag(0)\;,
\end{equation}
where
\begin{equation}
r=\cos(qt)-\frac{ik}{q}\sin(qt),s=-\frac{2ig_{\rm eff}}{q}\sin(qt)\;,
\end{equation}
with $k=2g_{\rm eff}+\omega_m$ and $q=\sqrt{k^2-4g_{\rm eff}^2}$. The uncertainty of the position quadrature $X$ is therefore given by
\begin{eqnarray}
\langle \Delta X(t)\rangle^2 & = & V\left[1-\frac{4g_{\rm eff}}{4g_{\rm eff}+\omega_m}\sin^2(qt)\right]\;,
\label{sq-osc}
\end{eqnarray}
where $V=\coth(\hbar\omega_m/2k_BT)$. This clearly represents a time-dependent squeezing, as $\langle\Delta X\rangle^2 \le \langle\Delta X\rangle^2_{g_{\rm eff}=0}=V$. At $qt=\pi/2$, the uncertainty becomes minimum as 
\begin{equation}
(\langle \Delta X(t)\rangle^2)_{\rm min}  =  V\frac{\omega_m}{4g_{\rm eff}+\omega_m}\;,
\end{equation}
that can further minimized by increasing $g_{\rm eff}$. From Eq. (\ref{sq-osc}), the relative squeezing can be expressed in decibel unit as
\begin{eqnarray}
S &=& -10\log_{10}\left[\langle\Delta X\rangle/\langle\Delta X\rangle_{g_{\rm eff}=0}\right]\nonumber\\
&=& -10\log_{10}\sqrt{1-\frac{4g_{\rm eff}}{4g_{\rm eff}+\omega_m}\sin^2(qt)}\;.
\end{eqnarray}
The relative squeezing $S$ thus does not depend upon the temperature.
The maximum squeezing
\begin{equation}
S_{\rm max}=5\log_{10}[4(g_{\rm eff}/\omega_m)+1]
\label{smax}
\end{equation}
occurs at $qt=\pi/2$ and it can be increased by increasing $g_{\rm eff}$. In Fig. 2, we show how the maximum attainable squeezing $S_{\rm max}$ varies with $g_{\rm eff}$ when the atom is prepared in one of the eigenstates.

It should be borne in mind that in the traditional models of quadratic coupling, the coupling strength $g_2$ between the cavity mode and the oscillator is much smaller than the linear coupling strength. This leads to the achievable squeezing only of the order of 1.8 dB \cite{shi-bhatta}. In the present case, we consider a hybrid model, which provides us a flexibility to increase the effective coupling to a much larger value, leading to larger squeezing. For example, for $g_{\rm eff}=1$  \cite{vacanti}, we have $S_{\rm max}=5\log 5\approx 3.5$ dB, beating the standard 50\% squeezing ($\equiv 3$ dB) limit for a bosonic system coupled to a thermal bath \cite{walls-book}. Further, this squeezing can be enhanced by moderately increasing $\epsilon$ (see Fig. 2). This effectively leads to the strong coupling limit, that would be useful to drive the mechanical oscillator and that can be achieved just by using large classical pump fields.
\begin{figure}[h]
\centering
\includegraphics[width=0.43\textwidth,
natwidth=600,natheight=410]{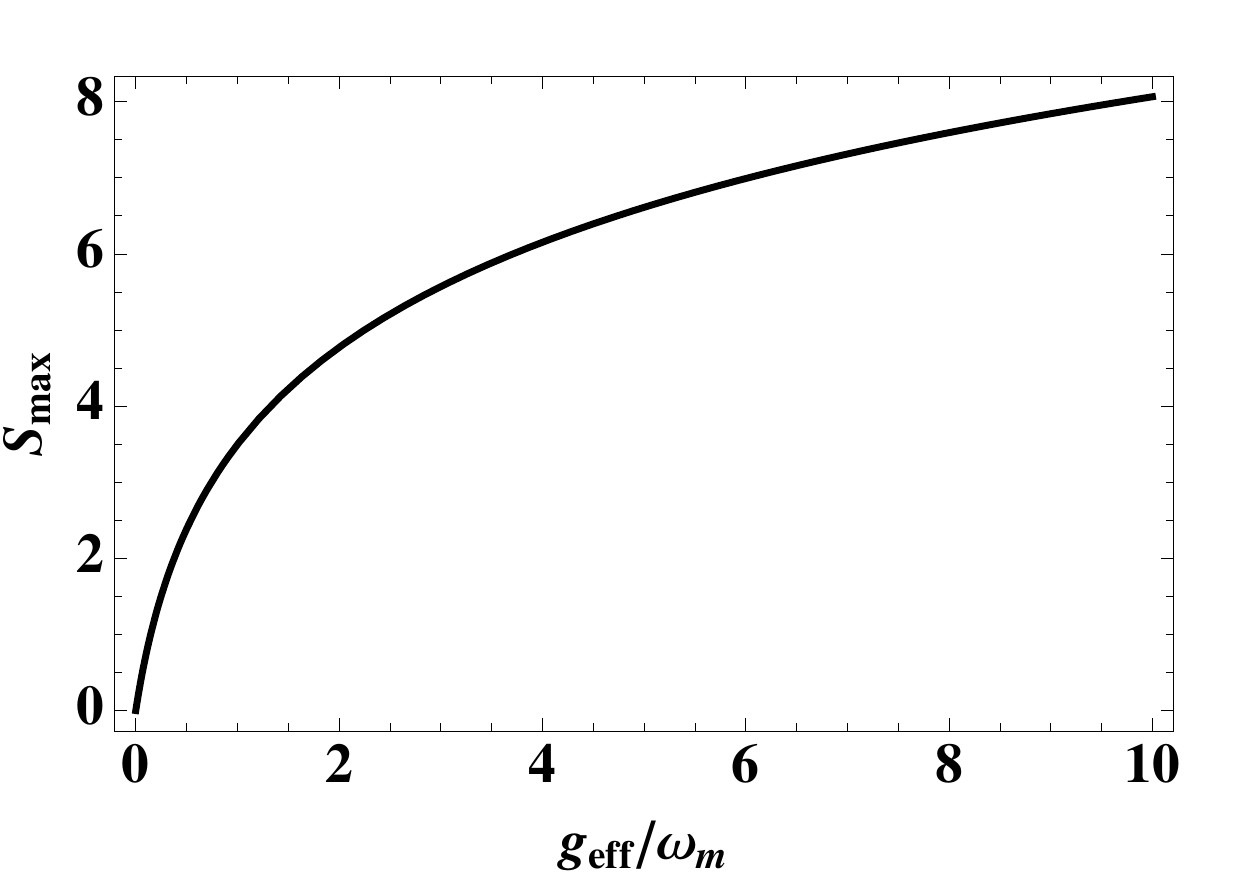}
\caption{Variation of maximum squeezing $S_{\rm max}$ (in decibel) with coupling constant $g_{\rm eff}/\omega_m$. }
\label{f:a1}
\end{figure}

\subsection{Squeezing spectrum}

As the evolution of the atom is effectively confined in the ground state manifold, the spontaneous emission may be ignored in the present study. Further the cavity mode, after its adiabatic elimination, does not significantly affect the squeezing through its decay. In the present model, the primary source of decoherence is the coupling of the mechanical oscillator to the thermal phononic bath at a temperature $T$, due to which the position uncertainty increases and becomes proportional to $V=\coth(\hbar\omega_m/k_BT)$ [see Eq. (\ref{sq-osc})].

The effect of decoherence of the mechanical oscillator can be further analyzed in terms of squeezing spectrum, where we introduce another annihilation operator $c(\omega)$ for the bosonic bath. The oscillator-bath interaction can be described by the following Hamiltonian \cite{orszag-book}:
\begin{eqnarray}
H_{tot}&=&H^{(4)}+H_{bath}+H_I\;,\nonumber\\
H_{bath}&=&\int d\omega\;\;\omega\;\; c^\dag(\omega)c(\omega)\nonumber\\
H_I&=&i\int d\omega K(\omega)[c^\dag(\omega)b-c(\omega)b^\dag]\;,
\end{eqnarray}
where $K(\omega)$ is the frequency-dependent coupling constant. In this case, the Heisenberg equation of motion of the mode $b$ can be written as
\begin{equation}
\dot{b}=-i[b,H^{(4)}]-\frac{\gamma}{2}b+\sqrt{\gamma}b_{I}(t)\;,
\label{b-eq}
\end{equation}
where we have chosen $K(\omega)=\sqrt{\gamma}$, as in the case of white bath and
\begin{equation}
b_{I}(t)=\int_{-\infty}^\infty d\omega \exp[-i\omega(t-t_0)]c(\omega, t_0)\;.
\end{equation}
The solution of Eq. (\ref{b-eq}) can be obtained in frequency domain, through the Fourier transform
\begin{equation}
b(\omega)=\frac{1}{2\pi}\int_{-\infty}^\infty e^{i\omega t}b(t)dt\;.
\end{equation}
We find that
\begin{equation}
b(\omega)=\frac{\sqrt{\gamma}\left[-2ig_{\rm eff} b_{I}^\dag(-\omega)-\{i(\omega+2g_{\rm eff}+\omega_m)-\frac{\gamma}{2}\}b_{I}(\omega)\right]}{\left(i\omega-\frac{\gamma}{2}\right)^2+\omega_m(4g_{\rm eff}+\omega_m)}\;.
\end{equation}
Therefore, by noting that $X(\omega)=[b(\omega)+b^\dag(\omega)]/2$, the position quadrature fluctuation $\langle X(\omega),X(\omega')\rangle=\langle X(\omega)X(\omega')\rangle-\langle X(\omega)\rangle\langle X(\omega')\rangle$ can be easily obtained. Using the following relations for the bath at thermal equilibrium at a temperature $T$
\begin{eqnarray}
\langle b_{I}^\dag(\omega)b_{I}(-\omega')\rangle &=&\bar{n}(\omega)\delta(\omega+\omega')\;,\nonumber\\
\langle b_{I}(\omega)b_{I}^\dag(-\omega')\rangle &=&[\bar{n}(\omega)+1]\delta(\omega+\omega')\;,\nonumber\\
\langle b_{I}(\omega)\rangle &=& \langle b_{I}^\dag(\omega)\rangle = 0\;,
\end{eqnarray}
we find that
\begin{eqnarray}
&&\langle X(\omega),X(\omega)\rangle = \frac{\gamma}{4}\frac{P}{Q}\;;\label{var}\\
P&=&(\bar{n}+1)\left\{\left(\frac{\gamma}{2}\right)^2+(\omega+2g_{\rm eff}+\omega_m)^2\right\}\nonumber\\
&&+\bar{n}\left\{\left(\frac{\gamma}{2}\right)^2+(\omega-2g_{\rm eff}-\omega_m)^2\right\}\nonumber\\
&&-2g_{\rm eff}\{\omega+(2\bar{n}+1)\omega_m\}\;,\nonumber\\
Q&=&\left[\left(\frac{\gamma}{2}\right)^2+\omega_m(4g_{\rm eff}+\omega_m)-\omega^2\}\right]^2+(\omega\gamma)^2\;.\nonumber
\end{eqnarray}
In Fig. 3, we display the spectrum of position uncertainty. It exhibits two maxima at the critical frequencies $\omega_{\rm crit}=\pm\sqrt{\omega_m(4g_{\rm eff}+\omega_m)-\gamma^2/4}$, where $Q$ becomes minimum.
Note that  the above variance (\ref{var}) decreases as $g_{\rm eff}$ increases, referring to squeezing at a particular frequency $\omega$ (see Fig. 4).  We also find that as the decay rate $\gamma$ of the oscillator increases, the uncertainty increases at $\omega=\omega_{\rm crit}$. This suggests that the decoherence degrades squeezing.
\begin{figure}[h]
\centering
\includegraphics[width=0.43\textwidth,
natwidth=600,natheight=200]{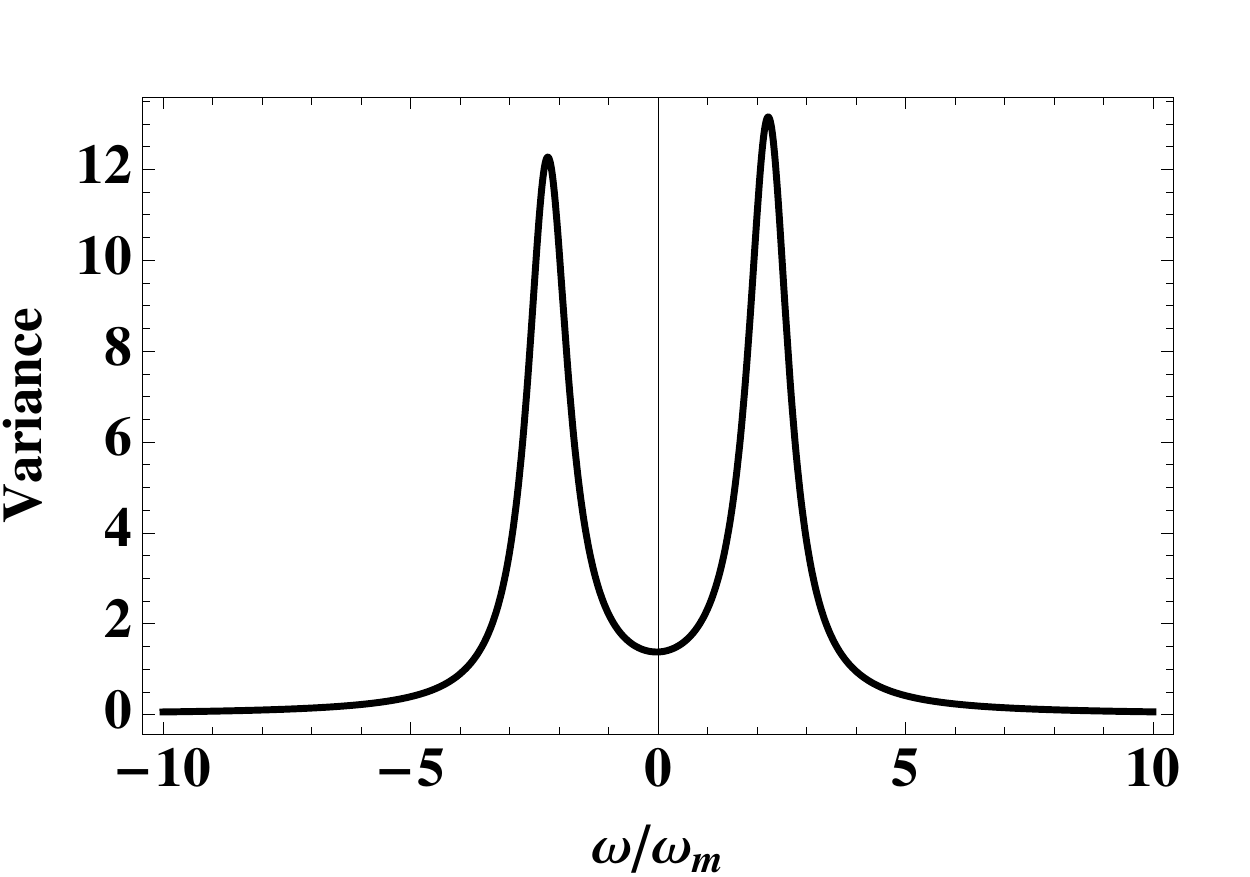}
\caption{Variation of the position uncertainty $\langle X(\omega),X(\omega)\rangle$ with the frequency $\omega/\omega_m$. We have chosen $\bar{n}=10$ and $\gamma=g_{\rm eff}=\omega_m$.}
\label{f:a11}
\end{figure}
\begin{figure}[h]
\centering
\includegraphics[width=0.43\textwidth,
natwidth=600,natheight=240]{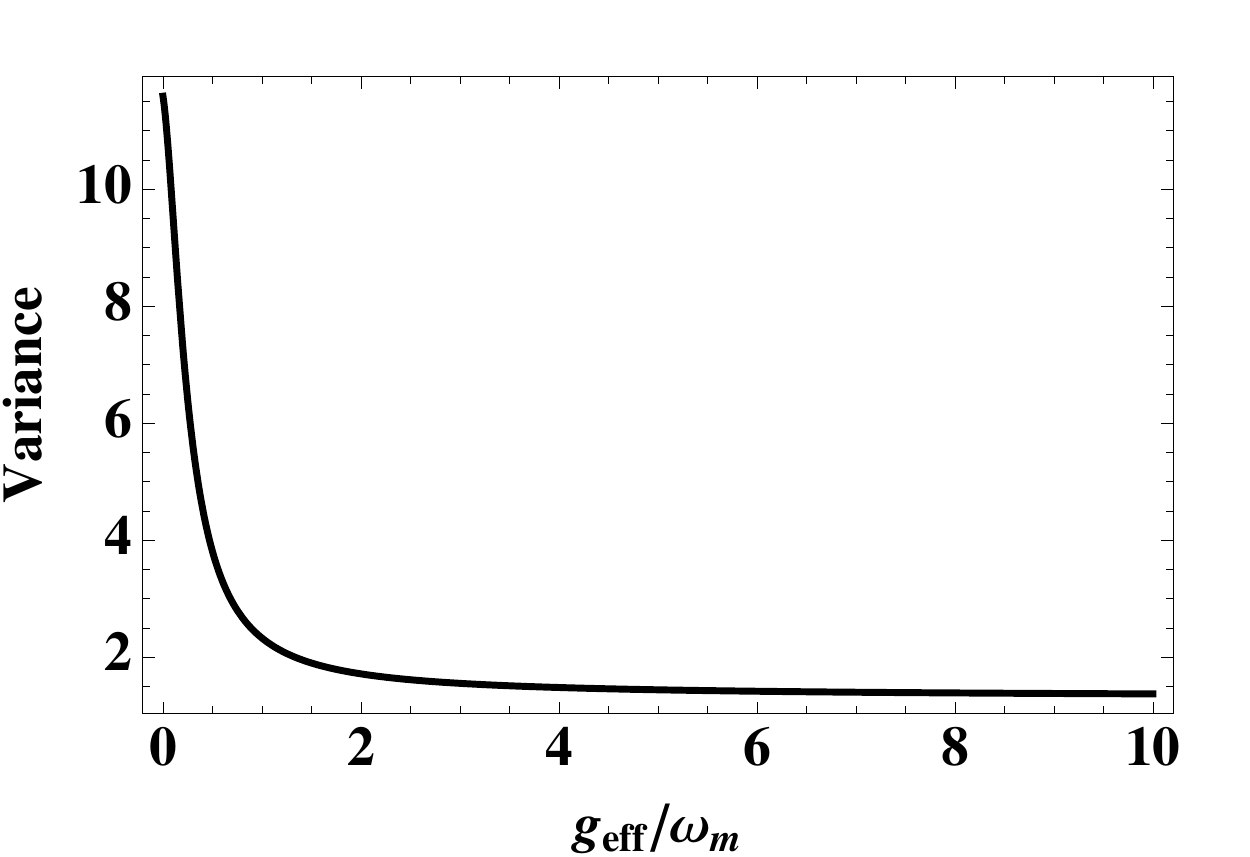}
\caption{Variation of the position uncertainty $\langle X(\omega),X(\omega)\rangle$ with the effective coupling constant $g_{\rm eff}/\omega_m$ at the frequency $\omega=\omega_m$. We have chosen $\bar{n}=10$ and $\gamma=\omega_m$.}
\label{f:a12}
\end{figure}

\section{Discussion}\label{s:iii}
As discussed above, the atom-assisted squeezing is neither affected by the decay of the atom nor by the cavity mode. In addition, the squeezing can be controlled dynamically using external classical fields $\Omega$ and $\epsilon$, used for driving the atomic transition and the cavity mode, respectively.  As clear from the Eq. (\ref{sq-osc}), the squeezing depends upon the effective atom-oscillator coupling constant $g_{\rm eff}=(g_2/\delta^2)(\lambda_i+\epsilon^2)$. Hence, for negligible cavity driving ($\epsilon \rightarrow 0$), the squeezing primarily depends upon the driving field $\Omega$. However in adiabatic limit, $\lambda_i \ll 1$ and therefore, the squeezing cannot be increased substantially to a large value.  On the other hand, for larger values of $\epsilon$, the squeezing can be increased as much as possible, as $g_{\rm eff}\approx (g_2/\delta^2)\epsilon^2$. For example, for $g_2/\delta\approx 10^{-2}$ and $\epsilon \sim 20$, one can have $g_{\rm eff}\approx 4$ and the maximum squeezing as large as 6 dB (see Fig. 2). We must emphasize that as $\epsilon$ increases, it does not substantially populate the cavity mode, as the cavity detuning $\delta$ is chosen to be large and therefore does not violate the condition of adiabatic elimination of cavity mode. We also note that an alternative way of achieving strong coupling regime could be to consider the microwave cavities and Rydberg atoms that have negligible decay rates \cite{haroche}. In this way, the coupling strength $g_2$ between the cavity mode and the mechanical mode could be made larger.

Squeezing  has also been considered by J\"ahne {\it et al.\/} \cite{jahne2009}, who had driven the cavity with a 8-10 dB squeezed light and thereafter transferred this squeezing to the mechanical oscillator to obtain 5 dB mechanical squeezing for strong coupling $g_{\rm eff}/\omega_m=0.1$. On the contrary, our technique does not rely upon such constraints. Just by pumping the cavity using a highly detuned field, one can achieve a squeezing as large as $>5$ dB, even in the weak coupling limit. Further, Asjad {\it et al.\/} \cite{asjad} had used a cavity, driven with a pulsed laser and obtained a squeezing $\sim$10 dB using open-loop feedback control, for $g_{\rm eff}/\omega_m = 10^{-8}$. This mechanism is however limited by requirement of high power short optical nanosecond pulses. In our case, a cw laser pump would suffice to achieve squeezing.
Girvin and coworkers \cite{nunnenkamp} had proposed to drive the cavity with two fields at different frequencies, but of equal strengths, while $g_{\rm eff}/\omega_m=0.1$. This may lead to certain squeezing; however, it is constrained to work in resolved side-band limit only ($\omega_m$ is much larger than the cavity decay rate). Our proposal does not require to work in this condition, as the cavity mode is adiabatically eliminated.

\section{Conclusion}\label{s:iv}
In conclusions, we consider a hybrid atom-optomechanical system with the membrane-in-the-middle setup.  An atom is trapped inside the cavity and dispersively interacts with the cavity mode, leading to squeezing in the position quadrature of the mechanical oscillator. We show that this squeezing is independent of spontaneous emission of the atom and the cavity decay. We also discuss how the squeezing depends upon the initial preparation of the atomic state. The squeezing can further be enhanced by increasing $g_{\rm eff}$, which can controlled externally by the classical fields that drive the atom and the cavity mode. As an example, we show that a squeezing of $S_{\rm max}=3.5 $ dB of the oscillator can be attained for a strong coupling $g_{\rm eff}=\omega_m$, that beats the standard 50\% squeezing (= 3 dB) limit.  We have also analytically derived the squeezing spectrum that exhibits two maxima, width of which increases by larger decay rate of the oscillator.

\end{document}